\documentclass[pra,twocolumn,showpacs,superscriptaddress,groupedaddress,floatfix]{revtex4-1}
\usepackage[english]{babel}
\usepackage{amsmath}
\usepackage{amsthm}
\usepackage{amssymb}
\usepackage{mathrsfs}
\usepackage{graphicx}  % needed for figures
\usepackage{dcolumn}   % needed for some tables
\usepackage{bm}        % for math
\usepackage{tikz}
\usepackage[margin=1in,footskip=0.25in]{geometry}

\begin{document}

\title{Magnetic forces in the absence of a classical magnetic field}

\author{I. L. Paiva,$^{1,2}$ Y. Aharonov,$^{1,2,3}$ J. Tollaksen,$^{1,2}$ and M. Waegell$^{2}$}
\affiliation{$^{1}$Schmid College of Science and Technology, Chapman University, Orange, California 92866, USA}
\affiliation{$^{2}$Institute for Quantum Studies, Chapman University, Orange, California 92866, USA}
\affiliation{$^{3}$School of Physics and Astronomy, Tel Aviv University, Tel Aviv 6997801, Israel}

\begin{abstract}
It is shown that, in some cases, the effect of discrete distributions of flux lines in quantum mechanics can be associated with the effect of continuous distributions of magnetic fields with special symmetries. In particular, flux lines with an arbitrary value of magnetic flux can be used to create energetic barriers, which can be used to confine quantum systems in specially designed configurations. This generalizes a previous work where such energy barriers arose from flux lines with half-integer fluxons. Furthermore, it is shown how the Landau levels can be obtained from a two-dimensional grid of flux lines. These results suggest that the classical magnetic force can be seen as emerging entirely from the Aharonov-Bohm effect. Finally, the basic elements of a semi-classical theory that models the emergence of classical magnetic forces from fields with special symmetries are introduced.
\end{abstract}

\maketitle

\section{Introduction}

The magnetic Aharonov-Bohm (AB) effect \citep{Aharonov1959} occurs when a quantum particle with charge $q$ encircles --- but does not enter --- a region with magnetic flux $\Phi_B$ on its interior. In classical physics, the dynamics of a point charge $q$ would not be affected by the presence of the magnetic field inside the region, but, in the quantum treatment, the charge accumulates a phase whose modular part corresponds to
\begin{equation}
\varphi_{AB} = \frac{q\Phi_B}{\hbar}. \label{AB effect}
\end{equation}
This effect has applications in many areas of physics \citep{Olariu1985, Peshkin1989, berry1989quantum, ford1994aharonov, vidal1998aharonov, tonomura2006aharonov, recher2007aharonov, russo2008observation, peng2010aharonov, bardarson2013quantum, noguchi2014aharonov, cohen2019geometric}, but some of its consequences and its implications on the foundations of quantum mechanics are still subjects of active discussion in the literature \cite{aharonov1994, keating2001force, aharonov2004effect, Aharonov2005, Aharonov2009, berry2010semifluxon, Kaufherr2011, Vaidman2012, kang2015locality, saldanha2016alternative, aharonov2016nonlocality}.

In this letter, we generalize Ref. \cite{paiva2019topological}, where we showed that an infinite lattice of solenoids acts as an energy barrier for quantum charges and that two solenoids can be used to confine low-energy charges in a sector of a long cavity. Even though the states of the ``trapped'' particles, in general, depend on the designed geometry, they would not exist if the AB effect did not hold. Because of this, we called these states \textit{topological bound states}. In that case, we assumed that the magnetic flux inside each solenoid was a half-integer fluxon. Here, however, we show that such a restriction is not necessary, and the results hold for an arbitrary value of magnetic flux.

Moreover, we discuss how this generalization allows us to construct a parallel between the forces associated with continuous and discrete distributions of magnetic fluxes. In particular, we discuss how Landau levels are obtained in a grid of flux lines. Because each flux line affects the dynamics of charges via the AB effect, its influence is restricted to quantum systems. However, we show that if, when taking the classical limit, the distance between flux lines in the grid goes to zero in an appropriate way, it is possible to build a semi-classical model where the force associated the grid can be seen as quantum counterparts of forces from classical continuous magnetic fields that are functions of two spatial coordinates (say, $x$ and $y$) in the direction of the third coordinate (say, $z$ axis), i.e., $\vec{B}=B(x,y)\hat{z}$.

\begin{center}
\begin{figure}
\includegraphics[width=\columnwidth]{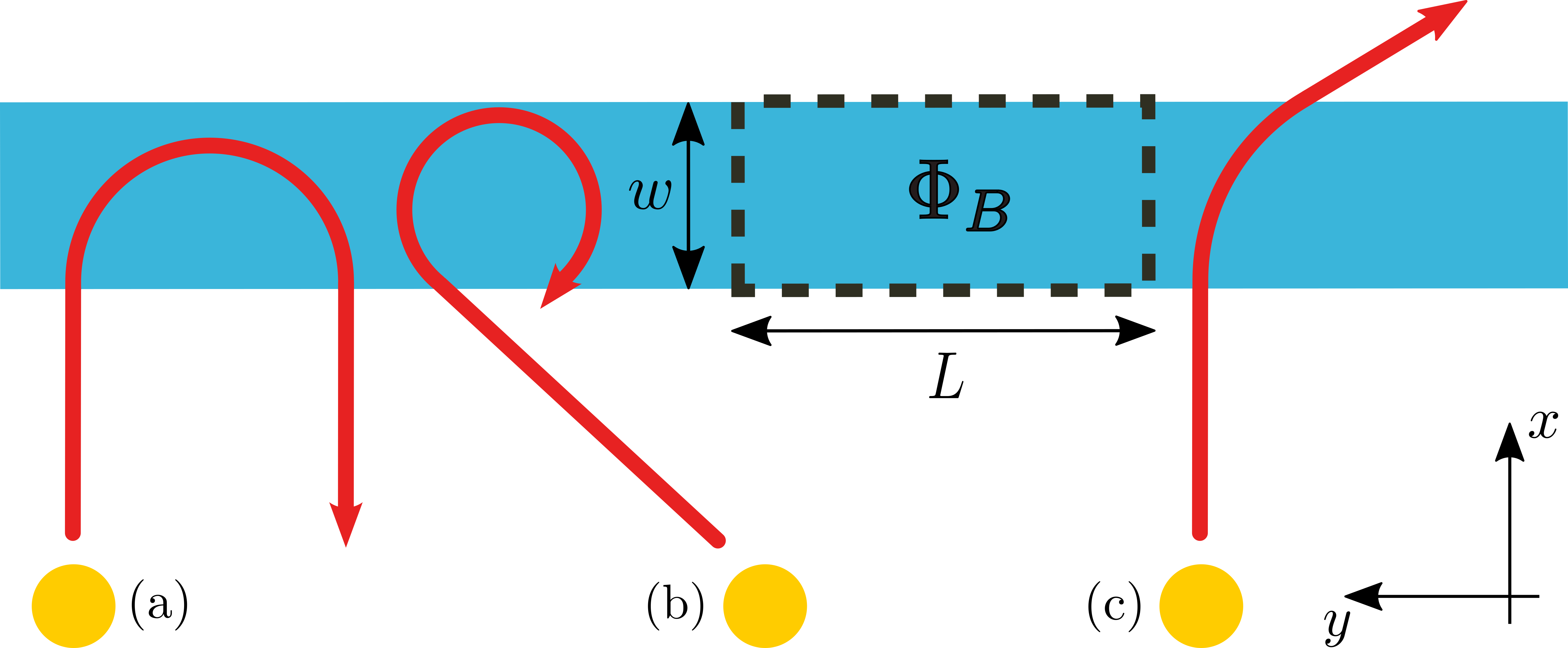}
\caption{Classical charges (yellow dots) traveling towards a wall of uniform magnetic field $\vec{B}=B\hat{z}$ (blue region) such that any rectangular region with length $L$ and width $w$ encloses a flux $\Phi_B=BLw$. Their trajectories are represented in red. In (a) and (b), low-energy particles cannot cross the wall, while a high-energy particle is scattered by it in (c). In general, if a particle passes through the wall, its transverse kinematic momentum is changed by $-q\Phi_B/L$.}
\label{fig1}
\end{figure}
\end{center}

\section{Walls of magnetic fields}

Let us start by considering the well-known effect in classical physics of having a (penetrable) wall parallel to the $y$ axis with a uniform magnetic field $\vec{B}=B\hat{z}$ and a point particle with mass $m$ and charge $q$ traveling in the $x$ direction, as represented in Fig. \ref{fig1}. Moreover, let $\Phi_B=BLw$ be the magnetic flux associated with any arbitrary region with length $L$ and width $w$. When the particle is inside the wall, it is deflected by the field into a circular arc of radius
\begin{equation}
R = \frac{mvw}{\left|q\phi_B\right|},
\label{radius}
\end{equation}
where $\phi_B=\Phi_B/L$ is the magnetic flux per unit of transverse length. Then, if the speed $v$ of the particle is such that $R<w$, i.e., if the particle's kinematic momentum $p_x=mv$ is such that
\begin{equation}
p_x<\left|q\phi_B\right|,
\label{classical-bound}
\end{equation}
the particle is reflected by the wall. Note that, here, and everywhere in the present letter, there is no bound on the $z$ component of the charge's kinematic momentum or energy, i.e., they can be arbitrarily large, and remain conserved without affecting our results.

Now, if we reduce the width $w$ of the wall, while simultaneously increasing the field magnitude $B$, in such a way that $\phi_B$ remains constant, then the maximum speed $v$ for which a given charge is reflected by the wall remains unchanged. This continues to hold even in the limit $w\rightarrow 0$, i.e., when the magnetic field wall becomes widthless. The barrier is, then, characterized entirely by the magnetic flux per unit of length $\phi_B$ associated with the wall.

Furthermore, Eq. \eqref{radius} is valid even if the particle is incident at an arbitrary angle in the $xy$ plane, as represented in Fig. \ref{fig1}(c). Then, in general, the wall reflects all those charges if $2R<w$, i.e., if their planar kinematic momentum $mv=\sqrt{p_x^2+p_y^2}$ is such that
\begin{equation}
\sqrt{p_x^2+p_y^2}<\frac{\left|q\phi_B\right|}{2}.
\label{general-cond}
\end{equation}

Finally, if the charge crosses the wall in a time interval $\Delta t$, the change in transverse kinematic momentum is
\begin{equation}
\Delta p_y = -qB\int_{\Delta t} v_x(t) dt = -qBw = -q\phi_B,
\end{equation}
i.e., the change in the transverse kinematic momentum is independent of the angle of incidence and the velocity of the particle. It is also straightforward to see that, when it is reflected, there is no change in transverse kinematic momentum of the charge after it leaves the wall, i.e., the angle of incidence equals the angle of reflection.

\begin{center}
\begin{figure}
\includegraphics[width=\columnwidth]{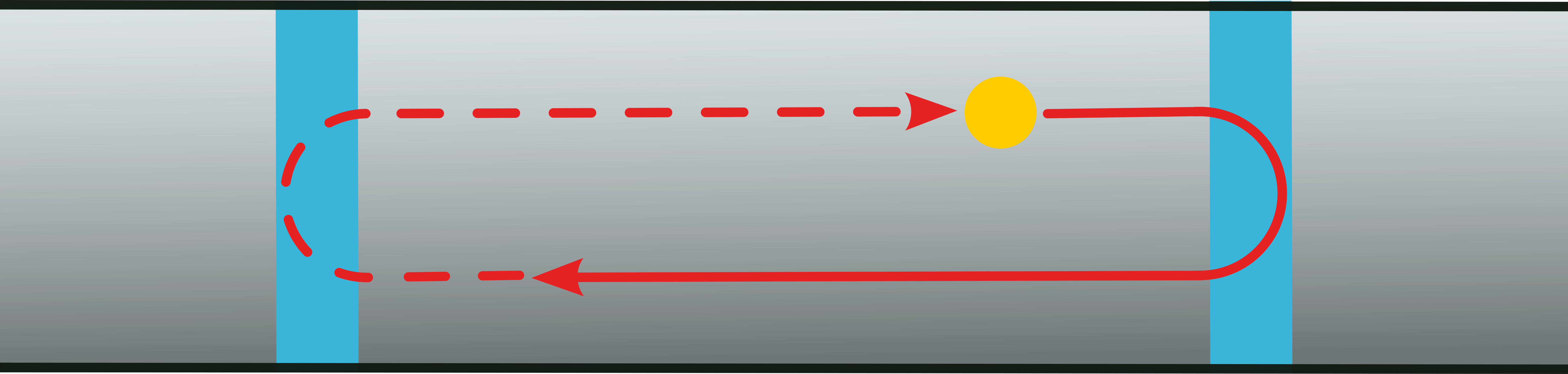}
\caption{Classical particle (yellow dot) trapped in a sector of a cavity by two walls of magnetic field (blue regions). The red curves represent the particle's trajectory.}
\label{fig2}
\end{figure}
\end{center}

Another example in classical physics that follows trivially from the above discussion is the scenario represented in Fig. \ref{fig2}, where two walls of magnetic field are placed inside a cavity. Clearly, each wall behaves like a barrier for low-energy particles, and there are bound states in the region between them. If the width of the walls is decreased to zero in a similar manner done in our previous example, only the flux per unit length associated with the wall determines the magnitude of the energy barrier it imposes.

We, now, analyze the direct analog of these results in quantum mechanics. In the first example, we considered an infinite widthless wall of magnetic field placed on the $y$ axis. A possible choice of vector potential for this scenario is $\vec{A}=\phi_B\Theta(x)\hat{y}$, where $\Theta$ is the step function. In this gauge, the Hamiltonian of the system can be written as
\begin{equation}
H = \frac{1}{2m} \left[P_x^2+\left(P_y-q\phi_B\Theta(X)\right)^2\right],
\label{hamiltonian}
\end{equation}
which implies that the initial average energy of the charge is
\begin{equation}
\langle E\rangle_i=\frac{\langle P_x^2\rangle_i+\langle P_y^2\rangle_i}{2m}
\end{equation}
and its final average energy, in case it crosses the lattice, is
\begin{equation}
\langle E\rangle_f=\frac{\langle P_x^2\rangle_f+\langle P_y^2\rangle_f+q^2\phi_B^2-2q\phi_B\langle P_y\rangle_f}{2m}.
\end{equation}
Because $\left[H,P_y\right]=0$, $\langle P_y\rangle_i=\langle P_y\rangle_f$ and $\langle P_y^2\rangle_i=\langle P_y^2\rangle_f$, i.e., canonical momentum is conserved. For comparison with the deflection in the classical case, note that the $y$ component of the kinematic momentum on the right-hand side is given by $mv_y = P_y - q\phi_B$ \footnote{To avoid possible confusion due to the asymmetry in $\vec{A}$, if the particle had started on the right-hand side with zero transverse kinetic momentum $mv_y$, its initial canonical momentum would be \unexpanded{$\langle P_y\rangle_i=q\phi_B$}. Since the expected value of \unexpanded{$P_y$} remains unchanged after the particle crosses the wall, it ends up with a transverse kinematic momentum of \unexpanded{$q\phi_B$}, which verifies the symmetry with the case of a particle incident from the left.}. Also, a charge cannot completely cross the wall \footnote{To be precise, higher-energy components of the wave packet pass through the wall, while lower-energy components are reflected.} whenever $\langle E\rangle_i<\langle E\rangle_f$. Then, if the particle is incident with average (kinematic) momentum $\langle P_x\rangle_i$ in the $x$ axis and $\langle P_y\rangle_i$ in the $y$ axis, its minimum average energy after crossing the wall is obtained when $\langle P_x\rangle_f\rightarrow0$, and it cannot completely pass it if
\begin{equation}
\langle P_x^2\rangle_i < q^2\phi_B^2-2q\phi_B\langle P_y\rangle_i.
\label{general-q-cond}
\end{equation}
It follows that, if the particle is perpendicularly incident, i.e., $\langle P_y\rangle_i=0$, the above condition becomes
\begin{equation}
\langle P_x\rangle_i < \left|q\phi_B\right|,
\label{q-cond1}
\end{equation}
where we used the fact that $\langle P_x^2\rangle\geq\langle P_x\rangle^2$. Observe that Eq. \eqref{q-cond1} is analogous to Eq. \eqref{classical-bound}. Moreover, if $\langle P_y\rangle_i\neq0$, Eq. \eqref{general-q-cond} only has a solution if its right-hand side is positive, i.e., if $\left|\langle P_y\rangle_i\right|<\left|q\phi_B\right|/2$. Then, the condition for a charge to be at least partially reflected by the wall, regardless of the angle of incidence, is
\begin{equation}
\sqrt{\langle P_x^2\rangle_i+\langle P_y^2\rangle_i} < \frac{\left|q\phi_B\right|}{2},
\end{equation}
which is analogous to Eq. \eqref{general-cond}. This shows an equivalence between the quantum and classical treatment of the problem.

Thus, as in the classical case, a cavity with two walls of magnetic field can be used to confine quantum charges with low energy. Consider a long cavity with width $L$. Furthermore, let the distance between the walls of magnetic field be $D$.

A particle that starts in the region between the walls must have at least an average energy of
\begin{equation}
\langle E \rangle = \frac{\pi^2\hbar^2}{2mL^2}+\frac{\pi^2\hbar^2}{2mD^2},
\label{initial-avg-energy}
\end{equation}
i.e., the minimum energy of a particle inside a two-dimensional box with side lengths $D$ and $L$. Moreover, the amount of average energy necessary for a charge to completely cross one of the walls is greater than or equal to
\begin{equation}
\langle E \rangle = \frac{\pi^2\hbar^2}{2mL^2}+\frac{q^2\phi_B^2}{2m}.
\label{final-avg-energy}
\end{equation}
Hence, if $\left|q\phi_B\right| > \pi\hbar/D$, i.e., if the separation $D$ between the walls is such that
\begin{equation}
D > \frac{\pi\hbar}{\left|q\phi_B\right|},
\label{bs-cond}
\end{equation}
there exist bound states created by the two walls inside the cavity. In the classical limit, i.e., when $\hbar\rightarrow0$, there is no restriction on the distance between the magnetic walls, as expected.

\section{Replacing walls by flux lines}

Now, we replace the widthless walls of magnetic field by flux lines --- or infinitely thin solenoids. In two dimensions, these lines are point objects. Without loss of generality, the magnetic flux on each line is assumed to be positive. In fact, it can be always achieved with a rotation of the referential system. Also, recall that the influence of each flux line in the dynamics of quantum charges is invariant under the addition of a fluxon, i.e., $\Phi_0 = 2\pi\hbar/q$. Because of this, we can consider magnetic fluxes limited to the interval $\left[0,\Phi_0\right)$.

\begin{center}
\begin{figure}
\includegraphics[width=\columnwidth]{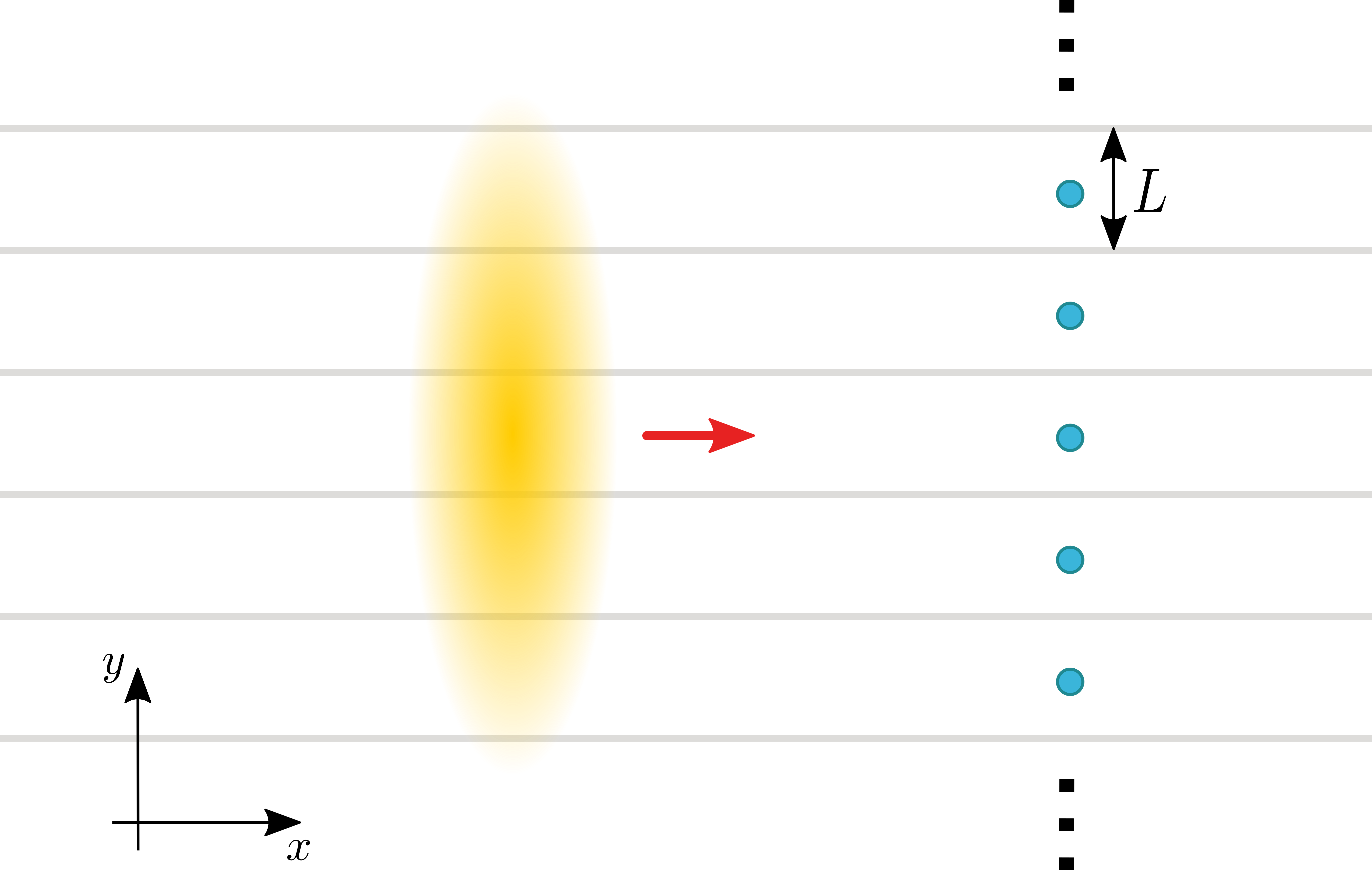}
\caption{Quantum charge $q$ (yellow cloud) sent through a lattice of solenoids (blue dots), each carrying a magnetic flux $\Phi_B$. Low-energy particles are reflected. Also, if the incident charges start with zeros on the gray lines, the diffraction due to the lattice can be used as a lower bound for particles inside cavities given by the regions between the gray lines.}
\label{fig3}
\end{figure}
\end{center}

Going back to the first scenario considered previously, we replace the continuous wall of magnetic field in free space with a lattice of flux lines with spacing $L$, with each line carrying a magnetic flux $\Phi_B$, as represented in Fig. \ref{fig3}. Classically, the particle's dynamics is no longer affected by the presence of the magnetic flux, since the quantum phase has no classical analog. However, in quantum mechanics, it is possible to show for an incident plane wave that, because of the scattering caused by the flux lines, the transverse kinematic momentum of the charge changes by \citep{Aharonov1969,Aharonov2005}
\begin{equation}
\Delta p_y = \frac{2\pi\hbar n}{L}-q\phi_B.
\label{scattering-momentum}
\end{equation}
That is, there exists a \textit{quantized} exchange of kinematic momentum between the lattice and the charge. Most importantly, there is a minimum change in the transverse kinematic momentum given by $\left|\Delta p_y\right|_{\text{min}}=\left|q\phi_B\right|$ if $\Phi_B\leq\Phi_0/2$, which is also to say that there is a minimum deflection as the charge passes through the lattice. If $\Phi_B>\Phi_0/2$, the minimum deflection is of $2\pi\hbar/L-q\phi_B$, which is the same as the minimum deflection if $-\Phi_0/2<\Phi_B<0$. Because of this extra symmetry, hereby we consider magnetic fluxes in the interval $\left[0,\Phi_0/2\right]$.

Let the charge start with average (kinematic) momentum $p_i=p_x^{(i)}\hat{x}+p_y^{(i)}\hat{y}$. Then, if $p_y^{(i)}=0$, the particle acquires a transverse kinematic momentum of at least $-q\phi_B$. Therefore, if $p_x^{(i)}<\left|q\phi_B\right|$ (Eq. \eqref{classical-bound}) is satisfied, the charge is reflected. Moreover, if $p_y^{(i)}\neq0$, its transverse kinematic momentum can, in principle, decrease in magnitude after it crosses the lattice. However, if $\left|p_y^{(i)}\right|<\left|\Delta p_y\right|_{\text{min}}/2$, the magnitude of the final transverse kinematic momentum of the particle cannot be smaller than $\left|\left|p_y^{(i)}\right|-\left|\Delta p_y\right|_{\text{min}}\right|$, which is still greater than $\left|p_y^{(i)}\right|$. In other words, if Eq. \eqref{general-cond} is satisfied, the charge must bounce off the lattice of flux lines. In conclusion, the lattice constitutes an energy barrier similar to the wall of magnetic field. Interestingly, Eq. \eqref{general-cond} does not depend explicitly on $\hbar$. However, because $\Phi_B$ is upper-bounded by $\Phi_0=2\pi\hbar/q$, for any fixed lattice spacing $L$, $\phi_B=\Phi_B/L\rightarrow0$ when $\hbar\rightarrow0$. However, $L$ can be adjusted so that $L\rightarrow0$ and $\phi_B$ is constant in that limit. This shows that, in specially designed configurations, consequences of the AB effect can still hold in the limit where $\hbar$ goes to zero.

\begin{center}
\begin{figure}
\includegraphics[width=\columnwidth]{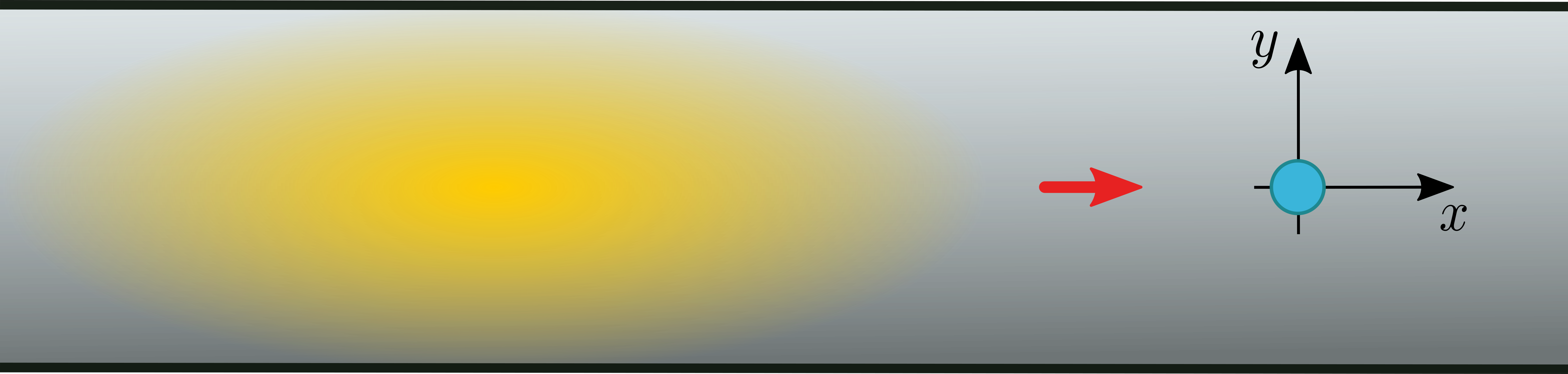}
\caption{Quantum charge $q$ (yellow cloud) inside a cavity traveling towards a flux line (blue dot) carrying a magnetic flux $\Phi_B$. Low-energy particles are reflected.}
\label{fig4}
\end{figure}
\end{center}

We now turn our attention to the second scenario considered previously, i.e., the cavity with two walls of magnetic field. Again, each wall is replaced by a single flux line carrying $\Phi_B$, as shown in Fig \ref{fig4}. Then, in classical physics, low-energy particles are no longer trapped in the region between the flux lines, but there are still quantum bound states. We present two arguments to corroborate this claim.

For our first argument, consider the cavity of Fig \ref{fig4} with a single flux line carrying a magnetic flux $\Phi_B$ placed at the origin of our system of reference, whose $x$ axis coincides with the long symmetry axis of the cavity. Assume the cavity has a long extension to the left and to the right of the flux line. Also, let the particle start in a separable state $\psi=\psi_x\psi_y$, where $\psi_y$ is the ground state in the transverse direction and $\psi_x$ is a state with low average energy in the direction of motion.

To treat the problem, we choose a gauge for which the vector potential $\vec{A}$ associated with the flux line is given by
\begin{equation}
\vec{A}(x,y) = \Phi_B \Theta(x) \delta(y) \hat{y}.
\label{vec-pot}
\end{equation}
Then, after crossing the flux line, $\psi_y$ turns into $\psi'_y$, which must have a phase discontinuity at $y=0$. Since $\psi_y$ already had the minimum amount of energy physically allowed, the discontinuity of $\psi'_y$ implies that its final average energy is greater than the initial energy of $\psi_y$. Therefore, if the initial average energy of $\psi_x$ is small enough, the charge cannot completely cross the flux line.

Since we cannot solve this case analytically, we begin by verifying this intuition with a perturbative analysis. Let the charge be prepared in a product state $\psi=\psi_x\psi_y$, where $\psi_y=\sqrt{2/L}\cos\left(\pi y/L\right)$ is the ground state in the $y$ direction. If the flux line in Fig. \ref{fig4} is initially carrying no magnetic flux, $\psi_y$ returns to its initial value after the packet has completely passed the flux line. Now, if the line carries a very small amount of magnetic flux $\epsilon\in\mathbb{R}$, then $\psi_y$ is barely disturbed as the charge passes it. This means that we can approximate the energy increase as
\begin{equation}
\begin{aligned}
\Delta\langle E\rangle &= \frac{1}{2m} \lim_{\gamma\rightarrow0} \int_{-\gamma}^\gamma \psi_y^* \left(P_y-q\epsilon\delta(y)\right)^2 \psi_y \ dy \\
                       &= \frac{1}{2m} \epsilon^2q^2 \lim_{\gamma\rightarrow0} \int_{-\gamma}^\gamma \delta(y)^2 \left|\psi_y\right|^2 \ dy.
\end{aligned}
\end{equation}
Because $\psi_y$ was already prepared with the minimum possible energy in the $y$ direction, this change of energy necessarily implies an increase in the average energy associated with that direction. We can conclude, then, that the flux line induces \textit{bound states} on its left (and on its right). As a result, as long as the initial average energy of $\psi_x$ is smaller than this threshold, the charge is at least partially reflected by the flux line.

This proves that flux lines carrying a magnetic flux $\Phi_B$ impose an energy barrier for charges. Our problem now concerns the quantification of the minimum amount of extra energy associated with $\psi'_y$ after the particle crosses the flux line. For that, we present our second argument.

Consider, once more, the lattice of flux lines represented in Fig. \ref{fig3}. Again, let the initial state of the charge be $\psi=\psi_x\psi_y$, where $\psi_y=\sqrt{2/L}\cos\left(\pi y/L\right)$, which has lines of zeros at $y=nL+L/2$, evenly spaced between the fluxes of the lattice, as shown in Fig. \ref{fig3}. For the incoming particle, the dynamics will be unchanged if infinite cavity walls (not magnetic) are added along the nodal lines, to the left of the lattice. Now, when the particle passes the flux lines, the walls end, and we get back to the case of free diffraction, where Eq. \eqref{scattering-momentum} applies. Then, adding in the walls on the right-hand side of the lattice to the initial Hamiltonian can only increase the minimum energy associated with the $y$ direction, and so the free case gives us a lower bound on the energy for the stacked cavity case. Hence, we conclude that the energy increase associated with $\psi_y$ after the charge crosses the flux line corresponds to at least the amount $q^2\phi_B^2/2m$. Moreover, following the same analysis, we can see that, for a general incident state, the minimum energy increase is $q^2\phi_B^2/8m$, just as in the classical and quantum wall cases. Thus, low-energy particles cannot cross the flux line.

With this in mind, we can consider the cavity with two flux lines separated by a distance $D$. If a charge starts in the region between the fluxes, the minimum amount of average energy it can have is given by Eq. \eqref{initial-avg-energy}. After crossing the flux line, the charge's minimum amount of average energy is expressed in Eq. \eqref{final-avg-energy}. As with the case of two walls of magnetic field inside a cavity, we conclude that, if Eq. \eqref{bs-cond} is satisfied, there exist bound states in the sector of the cavity delimited by the flux lines. Also, since such states exist because of the AB effect, we call them \textit{topological bound states}, as we did in Ref. \cite{paiva2019topological}.

\section{Discussion}

We have shown that the AB effect enables the construction of energy barriers with discrete distributions of magnetic field --- via the use of flux lines (solenoids). For magnetic fluxes between zero and $\Phi_0/2=\pi\hbar/q$, these barriers behave similarly to thin walls of continuous magnetic field. The similarity between the barriers implemented with walls of magnetic fields and with flux lines vanishes outside that interval because of the periodicity in the value of magnetic flux associated with the AB effect. However, this is not a significant limitation of our results. In fact, if a line of magnetic field has a flux $\Phi_B>\Phi_0/2$  associated to any $L$, there is a length $L'<L$ such that the magnetic flux $\Phi_B'$ associated with a region of that length is $\Phi_B'<\Phi_0$. Hence, in general, any widthless wall of uniform magnetic field with $\phi_B$ can be replaced by a lattice of flux lines with $\Phi_B=\phi_B L\leq\Phi_0/2=\pi\hbar/q$, where $L$ is the spacing of the lattice.

Also, one should notice that the region with magnetic field was taken to be widthless simply for convenience. In fact, our results imply that, in quantum mechanics, one can replace the effects of two-dimensional uniform magnetic fields with grids of flux lines. To see that, consider a region with constant magnetic field $\vec{B}=B\hat{z}$. Also, let this region be divided into squares with length $L$ such that each has a flux $\Phi_B=BL^2<\Phi_0/2$ --- or a magnetic flux per unit of transverse length $\phi_B=B/L$. Now, replace each square by a flux line with magnetic flux $\Phi_B$. Then, it is still possible to obtain the Landau levels with this two-dimensional square grid of flux lines with spacing $L$. In the singular gauge, the Hamiltonian of a charge can be written as
\begin{equation}
H = \frac{1}{2m} \left[P_x^2 + \left(P_y- q \sum_{n,s\in\mathbb{Z}} A_{ns}(X,Y) \right)^2\right], \label{hamiltonian-grid}
\end{equation}
where $A_{ns}(X,Y) = \Phi_B \Theta(X-nL) \delta(Y-sL)$ is the vector potential associated with each flux line. This Hamiltonian cannot be easily solved. However, it can be simplified by using the fact that, for each vertical layer, the average effect of the flux lines is a change of $q\phi_B$ in vertical momentum. Hence, the Hamiltonian in Eq. \eqref{hamiltonian-grid} can be approximated as
\begin{equation}
H = \frac{1}{2m} \left[P_x^2 + \left(P_y-q\phi_B \sum_{n,s\in\mathbb{Z}} \Theta(X-nL) \right)^2\right].
\end{equation}
Now, because the new expression for the Hamiltonian commutes with the canonical transverse momentum $P_y$, it is possible to replace this operator by its eigenvalue $\hbar k_y$. Then,
\begin{equation}
H = \frac{1}{2m} \left[P_x^2 + \left(\hbar k_y-q\phi_B \sum_{n,s\in\mathbb{Z}} \Theta(X-nL) \right)^2\right].
\end{equation}
One can easily see that the above Hamiltonian is formally an approximation of the one-dimensional harmonic oscillator. In fact, if $L\ll1$, the term $\phi_B \sum_{n,s\in\mathbb{Z}} \Theta(X-nL)$ can be approximated as $BX$. This shows that, indeed, the Landau levels can be recovered with the use of a two-dimensional grid of flux lines.

Moreover, recall that the magnetic flux $\Phi_B$ associated with a flux line vanishes in the classical limit. However, this does not necessarily imply that $\phi_B$ also vanishes. In fact, it is possible to take the distance $L$ between the flux lines to zero in the classical limit in a way that keeps $\phi_B$ constant. In this case, the minimum deflection does not vanish. It seems, then, that the AB effect generates a classical force. In fact, topological forces associated with the AB effect were previously discussed by Keating and Robin in Ref. \cite{keating2001force}, and even by Feynman in his well-known lectures \cite{feynman1989feynman} --- see, also, Refs. \cite{tiwari2018physical, pearle2018feynman} for a debate on Feynman's argument. However, the AB effect only occurs if the incident wave function is spread over enough lattice spacings --- and this spread has no analog for a classical particle.

\begin{center}
\begin{figure}
\includegraphics[width=\columnwidth]{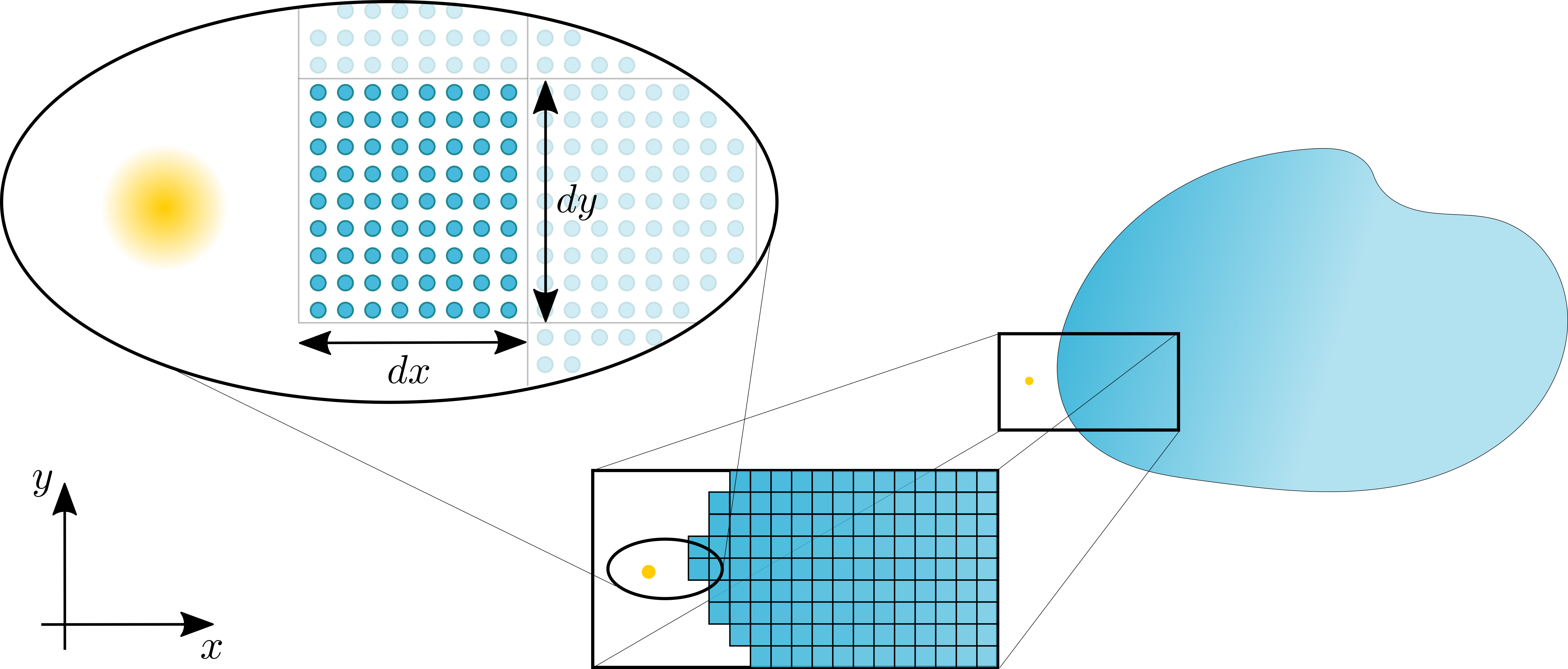}
\caption{Schematic representation of a semi-classical theory where a region with an arbitrary continuous distribution of magnetic field in the $z$ direction that does not depend on the $z$ coordinate (blue region) is replaced by a discrete distribution of flux lines (blue dots). The region can be split into infinitesimal areas, each with constant magnetic fields, as represented in the rectangular zoomed-in cut-away section. These infinitesimal areas can be replaced by a grid of flux lines, as shown in the elliptical zoomed-in cut-away section. The charge (yellow object) is assumed to have a spread much smaller than the infinitesimal areas.}
\label{semi-classical-fig}
\end{figure}
\end{center}

Nevertheless, under certain seemingly reasonable assumptions, the classical magnetic force in an arbitrary continuous field $\vec{B}(x,y) = B_z(x,y)\hat{z}$ can be seen as arising from the topological AB force. To show this, we first break the magnetic field up into differential squares of area $dx\cdot dy$, each with flux $d\Phi_B =  B_z(x,y) dx dy$, as illustrated in Fig. \ref{semi-classical-fig}. Then, we replace the uniform magnetic field $B_z$ in the differential region $dx\cdot dy$ by an $M\times N$ grid of differential point fluxes $d\Phi_B/MN$. Now, we consider a quantum charge $q$ spread over a region much smaller than $dx\cdot dy$ (but large enough to be diffracted by some of the flux lines) and incident on one of the infinitesimal cells with average velocity
\begin{equation}
\vec{v} = v_x \hat{x} + v_y \hat{y} + v_z\hat{z}.
\end{equation}
Also, we take the force as acting on the wave function of the particle when it is crossing the lattice. This assumption is consistent with a previous result where it was shown that there is a sudden change in the velocity distribution when the center of mass of the charge crosses a flux line \cite{aharonov2004effect}. Then, neglecting the $\hbar$ terms in Eq. \eqref{scattering-momentum}, which vanish in the classical limit, the change in kinematic momentum per vertical layer of $N$ fluxes in the $y$ direction amounts to $-q B_z(x,y) dx / M$. Similarly, the change in kinematic momentum in the $x$ direction per horizontal layer of $M$ flux lines is $q B_z(x,y) dy/N$. Then, if the center of the charge spread crossed $n_1\leq N$ vertical layers and $n_2\leq M$ horizontal layers while passing over the $dx\cdot dy$ infinitesimal cell, the total change in kinematic momentum can be approximated as
\begin{equation}
\begin{aligned}
d\vec{p} &= dp_x\hat{x}+dp_y\hat{y} \\
         &\approx qB_z(x,y) \frac{n_1}{M} dy \ \hat{x} - qB_z(x,y) \frac{n_2}{N} dx \ \hat{y}.
\end{aligned}
\end{equation}
Noticing that the particle's average velocity is kept approximately constant in each infinitesimal cell, i.e., $v_x\approx (n_2/N)(dx/dt)$ and $v_y\approx (n_1/M)(dy/dt)$, where $dt$ is the amount of time the center of the charge's distribution remains in the cell, a simple application of the chain rule gives
\begin{equation}
\vec{F} = \frac{dp_x}{dt} \hat{x} + \frac{dp_y}{dt} \hat{y} = qB_z v_y \hat{x} - qB_z v_x \hat{y},
\end{equation}
which correspond to
\begin{equation}
\vec{F} = q \vec{v} \times \vec{B}.
\end{equation}
Finally, taking the classical limit where the particle spread reduces to zero, $\vec{v}$ becomes the classical velocity, and $\hbar$ goes to zero, we obtain the classical force $\vec{F}$ experienced by a point charge $q$ in a magnetic field $\vec{B}$ --- using only the topological AB force.

This suggests that the AB effect in quantum mechanics may be the fundamental source of the classical magnetic force. It also lays the foundation for a semi-classical theory where the spread of the particle is introduced as a free parameter. We plan to fully develop this model in a future work.

Also, our results suggest experimental applications with the use of systems of solenoids where a simple manipulation of the current in each one serves as a control to emulate continuous magnetic fields with special symmetries.

Finally, we would like to mention that Refs. \cite{shelankov1998magnetic} and \cite{shelankov2000paraxial} came to our attention after this manuscript was finished. In those papers, Shelankov uses paraxial analysis to examine some of the same problems we have discussed here, restricted to wave functions of finite width. Our analysis uses straightforward Hamiltonian mechanics, and applies to general wave functions. We also examine topological bound states, which were absent from Shelankov's analysis.

\begin{acknowledgments}
This work was partially supported by the Fetzer Franklin Fund of the John E. Fetzer Memorial Trust. I.L.P. acknowledges financial support from the Science without Borders Program (CNPq/Brazil, Fund No. 234347/2014-7). Y.A. acknowledges support from the Israel Science Foundation (Grant 1311/14), Israeli Centers of Research Excellence (ICORE) Center ``Circle of Light,'' and DIP, the German-Israeli Project Cooperation.
\end{acknowledgments}

\bibliography{paper}
\bibliographystyle{apsrev4-1}
\end{document}